\documentclass[sigconf]{acmart}

\setcopyright{none}

\usepackage{times}
\usepackage[misc]{ifsym}
\usepackage{latexsym}
\usepackage{booktabs}
\usepackage{subcaption}
\usepackage{graphicx}
\usepackage{multirow}
\usepackage{amsmath}
\usepackage{xcolor}
\usepackage{comment}
\usepackage{footnote}
\usepackage{url}
\usepackage{xspace}
\usepackage{arydshln}
\newcommand{\PARADE}[1]{{PARADE}--{{#1}}\xspace}

\newcommand{\squishlist}{
 \begin{list}{$\bullet$}
  { \setlength{\itemsep}{0pt}
     \setlength{\parsep}{1pt}
     \setlength{\topsep}{1pt}
     \setlength{\partopsep}{0pt}
     \setlength{\leftmargin}{1.5em}
     \setlength{\labelwidth}{1em}
     \setlength{\labelsep}{0.5em} } }
\newcommand{\squishend}{
  \end{list}  }

\newcommand{\paragraphHdTop}[1] {\noindent\textbf{#1}} %
\newcommand{\paragraphHd}[1] {\vspace{3pt}\noindent\textbf{#1}}

\begin{document}

\title[PARADE: Passage Representation Aggregation for Document Reranking
]{PARADE: Passage Representation Aggregation\\ for Document Reranking
}

\newcommand{\an}[1]{$^{_{_{^{^{#1}}}}}$}

\author{Canjia Li\an{1,3*}, Andrew Yates\an{2}, Sean MacAvaney\an{4}, Ben He\an{1,3} and Yingfei Sun\an{1}}
\authornote{This work was conducted while the author was an intern at the Max Planck Institute for Informatics.}
\affiliation{\vspace{1ex}
	$^{1}$ University of Chinese Academy of Sciences, Beijing \country{China} \\ 
	$^{2}$ Max Planck Institute for Informatics, Saarbr\"ucken \country{Germany} \\
	$^{3}$ Institute of Software, Chinese Academy of Sciences Beijing \country{China}\\
    $^{4}$ IR Lab, Georgetown University, Washington, DC \country{USA}\\
    \texttt{licanjia17@mails.ucas.ac.cn, ayates@mpi-inf.mpg.de}\\
    \texttt{sean@ir.cs.georgetown.edu, \{benhe, yfsun\}@ucas.ac.cn} \\
    \vspace{5mm}
}

\renewcommand{\shortauthors}{Li et al.}

\begin{abstract}

Pretrained transformer models, such as BERT and T5, have shown to be highly effective at ad-hoc passage and document ranking. Due to inherent sequence length limits of these models, they need to be run over a document's passages, rather than processing the entire document sequence at once. Although several approaches for aggregating passage-level signals have been proposed, there has yet to be an extensive comparison of these techniques. In this work, we explore strategies for aggregating relevance signals from a document's passages into a final ranking score. We find that passage representation aggregation techniques can significantly improve over techniques proposed in prior work, such as taking the maximum passage score. We call this new approach PARADE.
In particular, PARADE can significantly improve results on collections with broad information needs where relevance signals can be spread throughout the document (such as TREC Robust04 and GOV2). Meanwhile, less complex aggregation techniques may work better on collections with an information need that can often be pinpointed to a single passage (such as TREC DL and TREC Genomics).
We also conduct efficiency analyses, and highlight several strategies for improving transformer-based aggregation.

\end{abstract}

\maketitle
\section{Introduction} \label{sec.introduction}
Pre-trained language models (PLMs), such as BERT~\cite{DBLP:conf/naacl/DevlinCLT19}, ELECTRA~\cite{DBLP:conf/iclr/ClarkLLM20} and T5~\cite{DBLP:journals/corr/abs-1910-10683}, have achieved state-of-the-art results on standard ad-hoc retrieval benchmarks.
The success of PLMs mainly relies on learning contextualized representations of input sequences using the transformer encoder architecture~\cite{DBLP:conf/nips/VaswaniSPUJGKP17}.
The transformer uses a self-attention mechanism whose computational complexity is quadratic with respect to the input sequence's length. Therefore, PLMs generally limit the sequence's length (e.g., to 512 tokens) to reduce computational costs.
Consequently, when applied to the ad-hoc ranking task, PLMs are commonly used to predict the relevance of passages or individual sentences~\cite{DBLP:conf/sigir/DaiC19,DBLP:conf/emnlp/YilmazWYZL19}.
The max or $k$-max passage scores (e.g., top 3) are then aggregated to produce a document relevance score.
Such approaches have achieved state-of-the-art results on a variety of ad-hoc retrieval benchmarks.

Documents are often much longer than a single passage, however, and intuitively there are many types of relevance signals that can only be observed in a full document.
For example, the {\it Verbosity Hypothesis}~\cite{DBLP:conf/sigir/RobertsonW94} states that relevant excerpts can appear at different positions in a document.
It is not necessarily possible to account for all such excerpts by considering only the top passages. %
Similarly, the ordering of passages itself may affect a document's relevance; a document with relevant information at the beginning is intuitively more useful than a document with the information at the end ~\cite{DBLP:conf/wsdm/HuiYBM18,Catena2019EnhancedNR}. %
Empirical studies support the importance of full-document signals.
\citeauthor{DBLP:conf/sigir/WuML0M19} study how passage-level relevance labels correspond to document-level labels, finding that more relevant documents also contain a higher number of relevant passages~\cite{DBLP:conf/sigir/WuML0M19}.
Additionally, experiments suggest that aggregating passage-level relevance scores to predict the document's relevance score outperforms the common practice of using the maximum passage score (e.g.,~\cite{DBLP:conf/ecir/BenderskyK08,DBLP:conf/sigir/FanGLXZC18,DBLP:conf/ecir/AiOC18}).

On the other hand, the amount of non-relevant information in a document can also be a signal, because relevant excerpts would make up a large fraction of an ideal document.
IR axioms encode this idea in the first length normalization constraint (LNC1), which states that adding non-relevant information to a document should decrease its score~\cite{10.1145/1961209.1961210}.
Considering a full document as input has the potential to incorporate signals like these.
Furthermore, from the perspective of training a supervised ranking model, the common practice of applying document-level relevance labels to individual passages is undesirable, because it introduces unnecessary noise into the training process.

In this work, we provide an extensive study on neural techniques for aggregating passage-level signals into document scores.
We study how PLMs like BERT and ELECTRA can be applied to the ad-hoc document ranking task while preserving many document-level signals. We move beyond simple passage \textit{score} aggregation strategies (such as Birch~\cite{DBLP:conf/emnlp/YilmazWYZL19}) and study passage \textit{representation} aggregation. We find that aggregation over passage representations using architectures like CNNs and transformers outperforms passage score aggregation.
Since the utilization of the full-text increases memory requirements, we investigate using knowledge distillation to create smaller, more efficient passage representation aggregation models that remain effective.
In summary, our contributions are: 

\begin{itemize}
\item The formalization of passage \textit{score} and \textit{representation} aggregation strategies, showing how they can be trained end-to-end, 
\item A thorough comparison of passage aggregation strategies on a variety of benchmark datasets, demonstrating the value of passage representation aggregation,
\item An analysis of how to reduce the computational cost of transformer-based representation aggregation by decreasing the model size,
\item An analysis of how the effectiveness of transformer-based representation aggregation is influenced by the number of passages considered, and
\item An analysis into dataset characteristics that can influence which aggregation strategies are most effective on certain benchmarks.
\end{itemize}

\begin{figure*}[tb]
    \centering
    \begin{subfigure}[t]{0.5\textwidth}
        \centering
        \includegraphics[height=1.2in, width=0.9\textwidth]{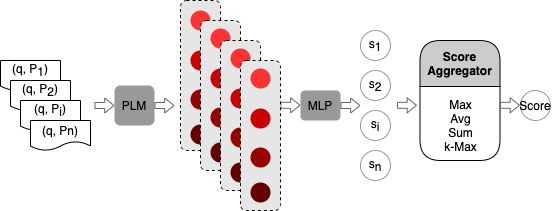}
        \caption{Previous approaches: score aggregation}
        \label{fig.comparison_score}
    \end{subfigure}%
    ~ 
    \begin{subfigure}[t]{0.5\textwidth}
        \centering
        \includegraphics[height=1.2in,width=0.9\textwidth]{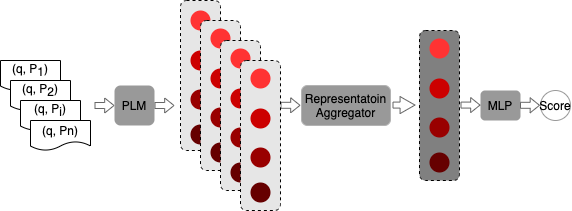}
        \caption{PARADE: representation aggregation}
        \label{fig.comparison_rep}
    \end{subfigure} %
    
    \caption{Comparison between score aggregation approaches  and PARADE's representation aggregation mechanism.}
    \label{fig.comparison_score_rep}
\vspace{8mm}
\end{figure*}

\begin{figure*}[tb]
    \centering
    \begin{subfigure}[t]{0.3\textwidth}
        \centering
        \includegraphics[height=1.2in]{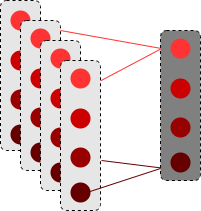}
        \caption{Max, Avg, Max, and Attn Aggregators}
        \label{fig.agg_mix}
    \end{subfigure}%
    ~ 
    \begin{subfigure}[t]{0.4\textwidth}
        \centering
        \includegraphics[height=1.2in]{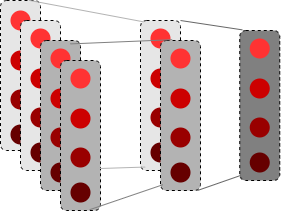}
        \caption{CNN Aggregator}
        \label{fig.agg_cnn}
    \end{subfigure} %
    ~
    \begin{subfigure}[t]{0.3\textwidth}
        \centering
        \includegraphics[height=1.2in]{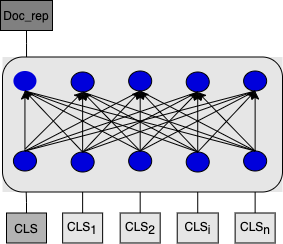}
        \caption{Transformer Aggregator}
        \label{fig.agg_trans}
    \end{subfigure} %
    \caption{Representation aggregators take passages' \texttt{[CLS]} representations as inputs and output a final document representation.}
    \label{fig.illustrated_rep}
\end{figure*}

\section{Related Work}
We review four lines of related research related to our study.

\paragraphHdTop{Contextualized Language Models for IR.}
Several neural ranking models have been proposed, such as 
DSSM~\cite{DBLP:conf/cikm/HuangHGDAH13}, 
DRMM~\cite{DBLP:conf/cikm/GuoFAC16}, \mbox{(Co-)PACRR}~\cite{DBLP:conf/emnlp/HuiYBM17,DBLP:conf/wsdm/HuiYBM18}, 
\mbox{(Conv-)KNRM}~\cite{DBLP:conf/sigir/XiongDCLP17,DBLP:conf/wsdm/DaiXC018}, and \mbox{TK}~\cite{Hofsttter2020InterpretableT}.
However, their contextual capacity is limited by relying on pre-trained unigram embeddings or using short n-gram windows. 
Benefiting from BERT's pre-trained contextual embeddings, BERT-based IR models have been shown to be superior to these prior neural IR models.
We briefly summarize related approaches here and refer the reader to a survey on transformers for text ranking by~\citet{lin2020pretrained} for further details.
These approaches use BERT as a relevance classifier in a cross-encoder configuration (i.e., BERT takes both a query and a document as input).
Nogueira et al. first adopted BERT to passage reranking tasks~\cite{DBLP:journals/corr/abs-1901-04085} using BERT's \texttt{[CLS]} vector.
Birch~\cite{DBLP:conf/emnlp/YilmazWYZL19} and BERT-MaxP~\cite{DBLP:conf/sigir/DaiC19} explore using sentence-level and passage-level relevance scores from BERT for document reranking, respectively.
CEDR proposed a joint approach that combines BERT’s outputs with existing neural IR models and handled passage aggregation via a representation aggregation technique (averaging)~\cite{DBLP:conf/sigir/MacAvaneyYCG19}.
In this work, we further explore techniques for passage aggregation and consider an improved CEDR variant as a baseline. We focus on the under-explored direction of representation aggregation by employing more sophisticated strategies, including using CNNs and transformers.

Other researchers trade off PLM effectiveness for efficiency by utilizing the PLM to improve document indexing~\cite{DBLP:journals/corr/abs-1904-08375, DBLP:journals/corr/abs-1910-10687}, pre-computing intermediate Transformer representations~\cite{Khattab2020ColBERTEA,MacAvaney2020EfficientDR,Gao2020EARLST,DBLP:conf/iclr/HumeauSLW20}, using the PLM to build sparse representations~\cite{MacAvaney2020ExpansionVP}, or reducing the number of Transformer layers~\cite{DBLP:journals/corr/abs-2002-01854,DBLP:conf/sigir/HofstatterZMCH20,DBLP:journals/corr/abs-2007-10434}.

Several works have recently investigated approaches for improving the Transformer's efficiency by reducing the computational complexity of its attention module, e.g., Sparse Transformer~\cite{DBLP:journals/corr/abs-1904-10509} and  Longformer~\cite{DBLP:journals/corr/abs-2004-05150}. 
QDS-Transformer tailors Longformer to the ranking task with query-directed sparse attention~\cite{DBLP:conf/emnlp/JiangXL020}.
We note that representation-based passage aggregation is more effective than increasing the input text size using the aforementioned models, but representation aggregation could be used in conjunction with such models.

\paragraphHd{Passage-based Document Retrieval.}
Callan first experimented with paragraph-based and window-based methods of defining passages~\cite{DBLP:conf/sigir/Callan94}.
Several works drive passage-based document retrieval in the language modeling context~\cite{DBLP:conf/cikm/LiuC02,DBLP:conf/ecir/BenderskyK08}, indexing context~\cite{DBLP:journals/bmcbi/Lin09}, and learning to rank context~\cite{DBLP:journals/ir/SheetritSK20}.
In the realm of neural networks, 
HiNT demonstrated that aggregating representations of passage level relevance can perform well in the context of pre-BERT models~\cite{DBLP:conf/sigir/FanGLXZC18}.
Others have investigated sophisticated evidence aggregation approaches~\cite{DBLP:conf/iclr/ZhaoXRSBT20,DBLP:conf/acl/ZhouHYLWLS19}.
Wu et al. explicitly modeled the importance of passages based on position decay, passage length, length with position decay, exact match, etc~\cite{DBLP:conf/sigir/WuML0M19}.
In a contemporaneous study, they proposed a model that considers passage-level representations of relevance in order to predict the passage-level cumulative gain of each passage~\cite{wuleveraging}.
In this approach the final passage's cumulative gain can be used as the document-level cumulative gain.
Our approaches share some similarities, but theirs differs in that they use passage-level labels to train their model and perform passage representation aggregation using a LSTM. 

\paragraphHdTop{Representation Aggregation Approaches for NLP.}
Representation learning has been shown to be powerful in many NLP tasks~\cite{DBLP:journals/pami/BengioCV13,DBLP:books/sp/LiuLS20}.
For pre-trained language models, a text representation is learned by feeding the PLM with a formatted text like \texttt{[CLS] TextA [SEP]} or \texttt{[CLS] TextA [SEP] TextB [SEP]}.
The vector representation of the prepended \texttt{[CLS]} token in the last layer is then regarded as either a text overall representation or a text relationship representation.
Such representations can also be aggregated for tasks that requires reasoning from multiple scopes of evidence.
Gear aggregates the claim-evidence representations by max aggregator, mean aggregator, or attention aggregator for fact checking~\cite{DBLP:conf/acl/ZhouHYLWLS19}.
Transformer-XH uses extra hop attention that bears not only in-sequence but also inter-sequence information sharing~\cite{DBLP:conf/iclr/ZhaoXRSBT20}.
The learned representation is then adopted for either question answering or fact verification tasks.
Several lines of work have explored hierarchical representations for document classification and summarization, including transformer-based approaches~\cite{yang2016hierarchical,liu2019hierarchical,zhang2019hibert}.
In the context of ranking, SMITH~\cite{DBLP:conf/cikm/Yang00BN20}, a long-to-long text matching model, learns a document representation with hierarchical sentence representation aggregation, which shares some similarities with our work.
Rather than learning independent document (and query) representations, SMITH is a bi-encoder approach that learns separate representations for each.
While such approaches have efficiency advantages, current bi-encoders do not match the effectiveness of cross-encoders, which are the focus of our work~\cite{lin2020pretrained}.

\paragraphHd{Knowledge Distillation.}
Knowledge distillation is the process of transferring knowledge from a large model to a smaller student model~\cite{DBLP:conf/nips/BaC14,DBLP:journals/corr/HintonVD15}.
Ideally, the student model performs well while consisting of fewer parameters.
One line of research investigates the use of specific distilling objectives for intermediate layers in the BERT model~\cite{DBLP:journals/corr/abs-1909-10351,DBLP:conf/emnlp/SunCGL19}, which is shown to be effective in the IR context~\cite{DBLP:journals/corr/abs-2009-07531}.
Turc et al. pre-train a family of compact BERT models and explore transferring task knowledge from large fine-tuned models~\cite{DBLP:journals/corr/abs-1908-08962}.
Tang et al. distill knowledge from the BERT model into Bi-LSTM~\cite{DBLP:journals/corr/abs-1903-12136}.
Tahami et al. propose a new cross-encoder architecture and transfer knowledge from this model to a bi-encoder model for fast retrieval~\cite{DBLP:journals/corr/abs-2004-11045}.
Hofstätter et al. also proposes a cross-architecture knowledge distillation framework using a Margin Mean Squared Error loss in a pairwise training manner~\cite{DBLP:journals/corr/abs-2010-02666}.
We demonstrate the approach in~\cite{DBLP:journals/corr/abs-1903-12136,DBLP:journals/corr/abs-2004-11045} can be applied to our proposed representation aggregation approach to improve efficiency without substantial reductions in effectiveness.

\section{Method}
In this section, we formalize approaches for aggregating passage representations into document ranking scores. We make the distinction between the passage \textit{score} aggregation techniques explored in prior work with passage \textit{representation} aggregation (PARADE) techniques, which have received less attention in the context of document ranking.
Given a query $q$ and a document $D$, a ranking method aims to generate a relevance score $rel(q, D)$ that estimates to what degree document $D$ satisfies the query $q$.
As described in the following sections, we perform this relevance estimation by aggregating passage-level relevance representations into a document-level representation, which is then used to produce a relevance score.

\subsection{Creating Passage Relevance Representations}
As introduced in Section~\ref{sec.introduction}, a long document cannot be considered directly by the BERT model\footnote{We refer to BERT since it is the most common PLM. In some of our later experiments, we consider the more recent and effective ELECTRA model~\cite{DBLP:conf/iclr/ClarkLLM20}; the same limitations apply to it and to most PLMs.} due to its fixed sequence length limitation.
As in prior work~\cite{DBLP:conf/sigir/DaiC19,DBLP:conf/sigir/Callan94}, we split a document into passages that can be handled by BERT individually.
To do so, a sliding window of 225 tokens is applied to the document with a stride of 200 tokens, formally expressed as $D=\{P_1, \ldots, P_n\}$ where $n$ is the number of passages.
Afterward, these passages are taken as input to the BERT model for relevance estimation.

Following prior work~\cite{DBLP:journals/corr/abs-1901-04085},
we concatenate a query $q$ and passage $P_i$ pair with a \texttt{[SEP]} token in between and another \texttt{[SEP]} token at the end. 
The special \texttt{[CLS]} token is also prepended, in which the corresponding output in the last layer is parameterized as a relevance representation $p_i^{cls} \in \mathcal{R}^d$, denoted as follows:

\begin{equation} \label{eq.cls_vec}
    p_{i}^{cls} = \operatorname{BERT}(q, P_i)
\end{equation}

\subsection{Score vs Representation Aggregation}
Previous approaches like BERT-MaxP~\cite{DBLP:conf/sigir/DaiC19} and Birch~\cite{DBLP:conf/emnlp/YilmazWYZL19} use a feedforward network to predict a relevance score from each passage representation $p_{i}^{cls}$, which are then aggregated into a document relevance score with a score aggregation approach.
Figure~\ref{fig.comparison_score} illustrates common score aggregation approaches like max pooling (``MaxP''), sum pooling, average pooling, and k-max pooling.
Unlike score aggregation approaches, our proposed representation aggregation approaches generate an overall document relevance representation by aggregating passage representations directly (see Figure~\ref{fig.comparison_rep}).
We describe the representation aggregators in the following sections.

\subsection{Aggregating Passage Representations}
Given the passage relevance representations $D^{cls}=\{p_1^{cls}, \ldots, p_n^{cls}\}$, PARADE summarizes $D^{cls}$ into a single dense representation $d^{cls}\in \mathcal{R}^d$ in one of several different ways, as illustrated in Figure~\ref{fig.illustrated_rep}.

{\bf \PARADE{Max}} utilizes a robust max pooling operation on the passage relevance features\footnote{Note that max pooling is performed on passage \textit{representations}, not over passage relevance scores as in prior work.}
in $D^{cls}$. As widely applied in Convolution Neural Network, max pooling has been shown to be effective in obtaining position-invariant features~\cite{DBLP:conf/icann/SchererMB10}.
Herein, each element at index $j$ in $d^{cls}$ is obtained by a element-wise max pooling operation on the passage relevance representations over the same index.

\begin{equation}\label{eq.PARADE_Max}
    d^{cls}[j] = \operatorname{max}  (p_{1}^{cls}[j], \ldots , p_{n}^{cls}[j])
\end{equation}

{\bf \PARADE{Attn}} assumes that each passage contributes differently to the relevance of a document to the query.
A simple yet effective way to learn the importance of a passage is to apply a feed-forward network to predict passage weights:

\begin{align}\label{eq.PARADE_Attn}
    w_1, \ldots, w_n & = \operatorname{softmax} (Wp_{1}^{cls}, \ldots, Wp_{n}^{cls}) \\
 d^{cls}& = \sum_{i=1}^n w_i p_{i}^{cls}
\end{align}
where $\operatorname{softmax}$ is the normalization function and $W \in \mathcal{R}^d$ is a learnable weight.

For completeness of study, we also introduce a {\bf \PARADE{Sum}} that simply sums the passage relevance representations.
This can be regarded as manually assigning equal weights to all passages (i.e., $w_i=1$).
We also introduce another {\bf \PARADE{Avg}} that is combined with document length normalization(i.e., $w_i=1/n$).

{\bf \PARADE{CNN}}, which operates in a hierarchical manner, stacks Convolutional Neural Network (CNN) layers with a window size of $d \times 2$ and a stride of 2.
In other words, the CNN filters operate on every pair of passage representations without overlap.
Specifically, we stack 4 layers of CNN, which halve the number of representations in each layer, as shown in Figure~\ref{fig.agg_cnn}.

{\bf \PARADE{Transformer}} enables passage relevance representations to interact by adopting the transformer encoder~\cite{DBLP:conf/nips/VaswaniSPUJGKP17} in a hierarchical way.
Specifically, BERT's \texttt{[CLS]}
token embedding and all $p_{i}^{cls}$ are concatenated, resulting in an input $x^l = (emb^{cls}, p_{1}^{cls}, \ldots , p_{n}^{cls})$ that is consumed by transformer layers to exploit the ordering of and dependencies among passages. That is,
\begin{align}\label{eq.PARADE_transformer}
    h = \operatorname{LayerNorm} (x^l + \operatorname{MultiHead} (x^l) \\
    x^{l+1} = \operatorname{LayerNorm} (h + \operatorname{FFN}(h))
\end{align}
where LayerNorm is the layer-wise normalization as introduced in~\cite{DBLP:journals/corr/BaKH16},
MultiHead is the multi-head self-attention~\cite{DBLP:conf/nips/VaswaniSPUJGKP17}, and
FFN is a two-layer feed-forward network with a ReLu activation in between.

As shown in Figure~\ref{fig.agg_trans}, the \texttt{[CLS]} vector of the last Transformer output layer, regarded as a pooled representation of the relevance between query and the whole document, is taken as $d^{cls}$.

\subsection{Generating the Relevance Score}
For all PARADE variants except \PARADE{CNN}, after obtaining the final $d^{cls}$ embedding, a single-layer feed-forward network (FFN) is adopted to generate a relevance score, as follows:
\begin{equation} \label{eq.pred}
    rel(q, D) = W_d d^{cls}
\end{equation}
where $W_d \in \mathcal{R}^d$ is a learnable weight.
For \PARADE{CNN}, a FFN with one hidden layer is applied to every CNN representation, and the final score is determined by the sum of those FFN output scores.

\subsection{Aggregation Complexity}

We note that the computational complexity of representation aggregation techniques are dominated by the passage processing itself. In the case of \PARADE{Max}, Attn, and Sum, the methods are inexpensive. 
For \PARADE{CNN} and \PARADE{Transformer}, there are inherently fewer passages in a document than total tokens, and (in practice) the aggregation network is shallower than the transformer used for passage modeling.

\section{Experiments}
\label{sec:experiment}

\begin{table}[tb]
    \centering
        \caption{Collection statistics. (There are 43 test queries in DL'19 and 45 test  queries in DL'20.)}
        \resizebox{.45\textwidth}{!}{
    \begin{tabular}{cccc} \toprule
    Collection & \# Queries & \# Documents & \# tokens / doc \\ \hline
      Robust04 & 249 & 0.5M & 0.7K \\
      GOV2     & 149 & 25M  & 3.8K \\ 
      Genomics & 64  & 162K    & 6.5K    \\ 
      MSMARCO    & 43/45  & 3.2M & 1.3K \\
     ClueWeb12-B13 & 80 & 52M   & 1.9K \\ \bottomrule
    \end{tabular}}

    \label{tab.stats}
\end{table}

\begin{table*}[tb] 
\centering
    \caption{Ranking effectiveness of PARADE on the {\it Robust04} and {\it GOV2} collection. 
    Best performance is in bold. 
    Significant difference between \PARADE{Transformer} and the corresponding method is marked with $\dagger$ ($p < 0.05$, two-tailed paired $t$-test). We also report the current best-performing model on Robust04 (T5-3B from~\cite{DBLP:journals/corr/abs-2003-06713}).}
    \resizebox{\textwidth}{!}{
\begin{tabular}{l|lll|lll|lll|lll} \toprule
& \multicolumn{3}{c}{Robust04 Title} & \multicolumn{3}{c}{Robust04 Description} & \multicolumn{3}{c}{GOV2 Title} & \multicolumn{3}{c}{GOV2 Description}\\ 
& MAP & P@20 & nDCG@20 & MAP & P@20 & nDCG@20 & MAP & P@20 & nDCG@20 & MAP & P@20 & nDCG@20 \\ \midrule 
BM25               & 0.2531$^{\dagger}$ &	0.3631$^{\dagger}$ &	0.4240$^{\dagger}$ &	0.2249$^{\dagger}$ &	0.3345$^{\dagger}$ &	0.4058$^{\dagger}$ & 0.3056$^{\dagger}$ & 0.5362$^{\dagger}$ & 	0.4774$^{\dagger}$ & 	0.2407$^{\dagger}$ & 	0.4705$^{\dagger}$ & 	0.4264$^{\dagger}$  \\ 
BM25+RM3           & 0.3033$^{\dagger}$ &	0.3974$^{\dagger}$ &	0.4514$^{\dagger}$ &	0.2875$^{\dagger}$ &	0.3659$^{\dagger}$ &	0.4307$^{\dagger}$ & 0.3350$^{\dagger}$ & 0.5634$^{\dagger}$ & 0.4851$^{\dagger}$ & 0.2702$^{\dagger}$ & 0.4993$^{\dagger}$ & 0.4219$^{\dagger}$   \\ \midrule
Birch              & 0.3763 & 0.4749$^{\dagger}$ & 0.5454$^{\dagger}$ & 0.4009$^{\dagger}$ & 0.5120$^{\dagger}$ & 0.5931$^{\dagger}$ & 0.3406$^{\dagger}$ & 0.6154$^{\dagger}$ & 0.5520$^{\dagger}$ & 0.3270 & 0.6312$^{\dagger}$ & 0.5763$^{\dagger}$\\ 
ELECTRA-MaxP       & 0.3183$^{\dagger}$ & 0.4337$^{\dagger}$ & 0.4959$^{\dagger}$ & 0.3464$^{\dagger}$ & 0.4731$^{\dagger}$ & 0.5540$^{\dagger}$ & 0.3193$^{\dagger}$ & 0.5802$^{\dagger}$ & 0.5265$^{\dagger}$ & 0.2857$^{\dagger}$ & 0.5872$^{\dagger}$ & 0.5319$^{\dagger}$\\
T5-3B (from~\cite{DBLP:journals/corr/abs-2003-06713})&-&-&-&0.4062&-&0.6122&-&-&-&-&-&-\\
ELECTRA-KNRM       & 0.3673$^{\dagger}$ & 0.4755$^{\dagger}$ & 0.5470$^{\dagger}$ & 0.4066 & {\bf 0.5255} & 0.6113 & 0.3469$^{\dagger}$ & 0.6342$^{\dagger}$ & 0.5750$^{\dagger}$ & 0.3269 & 0.6466 & 0.5864$^{\dagger}$\\
CEDR-KNRM (Max)    & 0.3701$^{\dagger}$ & 0.4769$^{\dagger}$ & 0.5475$^{\dagger}$ & 0.3975$^{\dagger}$ & 0.5219 & 0.6044$^{\dagger}$ & 0.3481$^{\dagger}$ & 0.6332$^{\dagger}$ & 0.5773$^{\dagger}$ & {\bf 0.3354}$^{\dagger}$ & 0.6648 & 0.6086 \\
\midrule
PARADE-Avg         & 0.3352$^{\dagger}$ & 0.4464$^{\dagger}$ & 0.5124$^{\dagger}$ & 0.3640$^{\dagger}$ & 0.4896$^{\dagger}$ & 0.5642$^{\dagger}$ & 0.3174$^{\dagger}$ & 0.6225$^{\dagger}$ & 0.5741$^{\dagger}$ & 0.2924$^{\dagger}$ & 0.6228$^{\dagger}$ & 0.5710$^{\dagger}$\\
PARADE-Sum         & 0.3526$^{\dagger}$ & 0.4711$^{\dagger}$ & 0.5385$^{\dagger}$ & 0.3789$^{\dagger}$ & 0.5100$^{\dagger}$ & 0.5878$^{\dagger}$ & 0.3268$^{\dagger}$ & 0.6218$^{\dagger}$ & 0.5747$^{\dagger}$ & 0.3075$^{\dagger}$ & 0.6436$^{\dagger}$ & 0.5879$^{\dagger}$ \\
PARADE-Max         & 0.3711$^{\dagger}$ & 0.4723$^{\dagger}$ & 0.5442$^{\dagger}$ & 0.3992$^{\dagger}$ & 0.5217 & 0.6022 & 0.3352$^{\dagger}$ & 0.6228$^{\dagger}$ & 0.5636$^{\dagger}$ & 0.3160$^{\dagger}$ & 0.6275$^{\dagger}$ & 0.5732$^{\dagger}$\\
PARADE-Attn        & 0.3462$^{\dagger}$ & 0.4576$^{\dagger}$ & 0.5266$^{\dagger}$ & 0.3797$^{\dagger}$ & 0.5068$^{\dagger}$ & 0.5871$^{\dagger}$ & 0.3306$^{\dagger}$ & 0.6359$^{\dagger}$ & 0.5864$^{\dagger}$ & 0.3116$^{\dagger}$ & 0.6584 & 0.5990\\
PARADE-CNN         & {\bf 0.3807 } & 0.4821$^{\dagger}$ & 0.5625 & 0.4005$^{\dagger}$ & 0.5249 & 0.6102 & 0.3555$^{\dagger}$ & 0.6530 & 0.6045 & 0.3308 & {\bf 0.6688} & {\bf 0.6169}\\
PARADE-Transformer & 0.3803 & {\bf 0.4920} & {\bf 0.5659} & {\bf 0.4084} & {\bf 0.5255} & {\bf 0.6127} & {\bf 0.3628} & {\bf 0.6651} & {\bf 0.6093} & 0.3269 & 0.6621 & 0.6069 \\ \bottomrule
\end{tabular}}
\label{tb.main_result}
\end{table*}

\subsection{Datasets} \label{sec:PARADE_dataset}

We experiment with several ad-hoc ranking collections.
Robust04\footnote{\url{https://trec.nist.gov/data/qa/T8_QAdata/disks4_5.html}} is a newswire collection used by the TREC 2004 Robust track.
GOV2\footnote{\url{http://ir.dcs.gla.ac.uk/test_collections/gov2-summary.htm}} is a web collection crawled from US government websites used in the TREC Terabyte 2004--06 tracks.
For Robust04 and GOV2, we consider both keyword (title) queries and description queries in our experiments.
The Genomics dataset~\cite{Hersh2006TrecGenomics,Hersh2007TrecGenomics} consists of scientific articles from the Highwire Press\footnote{\url{https://www.highwirepress.com/}} with natural-language queries about specific genes, and was used in the TREC Genomics 2006--07 track.
The MSMARCO document ranking dataset\footnote{\url{https://microsoft.github.io/TREC-2019-Deep-Learning}} is a large-scale collection and is used in TREC 2019--20 Deep Learning Tracks~\cite{DBLP:conf/trec/CraswellMYCV19,DBLP:conf/trec/CraswellMYC20}.
To create document labels for the development and training sets, passage-level labels from the MSMARCO passage dataset are transferred to the corresponding source document that contained the passage.
In other words, a document is considered relevant as long as it contains a relevant passage, and each query can be satisfied by a single passage.
The ClueWeb12-B13 dataset\footnote{\url{http://lemurproject.org/clueweb12/}} is a large-scale collection crawled from the web between February 10, 2012 and May 10, 2012. 
It is used for the NTCIR We Want Web 3 (WWW-3) Track~\cite{WWW-3}.
The statistics of these datasets are shown in Table~\ref{tab.stats}.
Note that the average document length is obtained only from the documents returned by BM25. 
Documents in GOV2 and Genomics are much longer than Robust04, making it more challenging to train an end-to-end ranker.

\subsection{Baselines}
We compare PARADE against the following traditional and neural baselines, including those that employ other passage aggregation techniques.

{\bf BM25} 
is  an  unsupervised  ranking  model  based  on  IDF-weighted counting~\cite{DBLP:conf/trec/RobertsonWHGP95}.
The documents retrieved by BM25 also serve as the candidate documents used with reranking methods.

{\bf BM25+RM3} is a query expansion model based on RM3~\cite{DBLP:conf/sigir/LavrenkoC01}.
We used Anserini's~\cite{DBLP:journals/jdiq/YangFL18} implementations of BM25 and BM25+RM3.
Documents are indexed and retrieved with the default settings for keywords queries.
For description queries, we set $b=0.6$ and changed the number of expansion terms to 20.

{\bf Birch} 
aggregates sentence-level evidence provided by BERT to rank documents~\cite{DBLP:conf/emnlp/YilmazWYZL19}.
Rather than using the original Birch model provided by the authors, we train an improved ``Birch-Passage'' variant.
Unlike the original model, Birch-Passage uses passages rather than sentences as input, it is trained end-to-end, it is fine-tuned on the target corpus rather than being applied zero-shot, and it does not interpolate retrieval scores with the first-stage retrieval method.
These changes bring our Birch variant into line with the other models and baselines (e.g., using passages inputs and no interpolating), and they additionally improved effectiveness over the original Birch model in our pilot experiments.

{\bf ELECTRA-MaxP}
adopts the maximum score of passages within a document as an overall relevance score~\cite{DBLP:conf/sigir/DaiC19}.
However, rather than fine-tuning BERT-base on a Bing search log, we improve performance by fine-tuning on the MSMARCO passage ranking dataset.
We also use the more recent and efficient pre-trained ELECTRA model rather than BERT.

{\bf ELECTRA-KNRM} 
is a kernel-pooling neural ranking model based on query-document similarity matrix~\cite{DBLP:conf/sigir/XiongDCLP17}.
We set the kernel size as 11.
Different from the original work, we use the embeddings from the pre-trained ELECTRA model for model initialization.

{\bf CEDR-KNRM (Max)}
combines the advantages from both KNRM and pre-trained model~\cite{DBLP:conf/sigir/MacAvaneyYCG19}.
It digests the kernel features learned from KNRM and the \texttt{[CLS]} representation as ranking feature. We again replace the BERT model with the more effective ELECTRA.
We also use a more effective variant that performs max-pooling on the passages' \texttt{[CLS]} representations, rather than averaging.

{\bf T5-3B}
defines text ranking in a sequence-to-sequence generation context using the pre-trained T5 model~\cite{DBLP:journals/corr/abs-2003-06713}.
For document reranking task, it utilizes the same score max-pooling technique as in BERT-MaxP~\cite{DBLP:conf/sigir/DaiC19}. Due to its large size and expensive training, we present the values reported by~\cite{DBLP:journals/corr/abs-2003-06713} in their zero-shot setting, rather than training it ourselves.

\subsection{Training} \label{sec:training}
To prepare the ELECTRA model for the ranking task, we first fine-tune ELECTRA on the MSMARCO passage ranking dataset~\cite{msmarco}.
The fine-tuned ELECTRA model is then used to initialize PARADE's PLM component.
For \PARADE{Transformer} we use two randomly initialized transformer encoder layers with the same hyperparemeters  (e.g., number of attention heads, hidden size, etc.) used by BERT-base.
Training of PARADE and the baselines was performed on a single Google TPU v3-8 using a pairwise hinge loss.
We use the Tensorflow implementation of PARADE available in the Capreolus toolkit~\cite{yates2020capreolus}, and a standalone imiplementation is also available\footnote{\url{https://github.com/canjiali/PARADE/}}.
We train on the top 1,000 documents returned by a first-stage retrieval method; documents that are labeled relevant in the ground-truth are taken as positive samples and all other documents serve as negative samples.
We use BM25+RM3 for first-stage retrieval on Robust04 and BM25 on the other datasets with parameters tuned on the dev sets via grid search.
We train for 36 ``epochs'' consisting of 4,096 pairs of training examples with a learning rate of 3e-6, warm-up over the first ten epochs, and a linear decay rate of 0.1 after the warm-up.
Due to its larger memory requirements, we use a batch size of 16 with CEDR and a batch size of 24 with all other methods.
Each instance comprises a query and all split passages in a document.
We use a learning rate of 3e-6 with warm-up over the first 10 proportions of training steps.

Documents are split into a maximum of 16 passages. 
As we split the documents using a sliding window of 225 tokens with a stride of 200 tokens, a maximum number of 3,250 tokens in each document are retained.
The maximum passage sequence length is set as 256.
Documents with fewer than the maximum number of passages are padded and later masked out by passage level masks.
For documents longer than required, the first and last passages are always kept while the remaining are uniformly sampled as in~\cite{DBLP:conf/sigir/DaiC19}.

\begin{table}[tb]
\centering
    \caption{Ranking effectiveness on the {\it Genomics} collection. Significant difference between \PARADE{Transformer} and the corresponding method is marked with $\dagger$ ($p < 0.05$, two-tailed paired $t$-test). The top neural results are listed in bold, and the top overall scores are underlined.}
    \label{tab:genomics}
\begin{tabular}{l|lll} \toprule
& MAP & P@20 & nDCG@20  \\ \midrule
BM25                & 0.3108 & 0.3867 & 0.4740 \\
TREC Best           & \underline{0.3770} & \underline{0.4461} & \underline{0.5810} \\ \midrule
Birch               & 0.2832 & 0.3711 & 0.4601 \\
BioBERT-MaxP        & 0.2577 & 0.3469 & 0.4195$^{\dagger}$ \\
BioBERT-KNRM        & 0.2724 & 0.3859 & 0.4605 \\
CEDR-KNRM (Max)     & 0.2486 & 0.3516$^{\dagger}$ & 0.4290 \\
\midrule
PARADE-Avg          & 0.2514$^{\dagger}$ & 0.3602 & 0.4381 \\
PARADE-Sum          & 0.2579$^{\dagger}$ & 0.3680 & 0.4483 \\
PARADE-Max          &\bf0.2972 &\bf0.4062$^{\dagger}$ & \bf 0.4902 \\
PARADE-Attn         & 0.2536$^{\dagger}$ & 0.3703 & 0.4468 \\
PARADE-CNN          & 0.2803 & 0.3820 & 0.4625 \\
PARADE-Transformer  & 0.2855 & 0.3734 & 0.4652 \\ \bottomrule
\end{tabular}
\end{table}

\subsection{Evaluation}
Following prior work~\cite{DBLP:conf/sigir/DaiC19, DBLP:conf/sigir/MacAvaneyYCG19}, we use 5-fold cross-validation.
We set the reranking threshold to 1000 on the test fold as trade-off between latency and effectiveness.
The reported results are based on the average of all test folds.
Performance is measured in terms of the MAP, Precision, ERR and nDCG ranking metrics using \texttt{trec\_eval}\footnote{\url{https://trec.nist.gov/trec_eval}} with different cutoff.
For NTCIR WWW-3, the results are reported using \texttt{NTCIREVAL} \footnote{\url{http://research.nii.ac.jp/ntcir/tools/ntcireval-en.html}}.

\subsection{Main Results}
\label{sec:results}

The reranking effectiveness of PARADE on the two commonly-used Robust04 and GOV2 collections is shown in Table~\ref{tb.main_result}.
Considering the three approaches that do not introduce any new weights, \PARADE{Max} is usually more effective than \PARADE{Avg} and \PARADE{Sum}, though the results are mixed on GOV2.
\PARADE{Max} is consistently better than \PARADE{Attn} on Robust04, but \PARADE{Attn} sometimes outperforms \PARADE{Max} on GOV2.
The two variants that consume passage representations in a hierarchical manner, \PARADE{CNN} and \PARADE{Transformer}, consistently outperforms the four other variants.
This confirms the effectiveness of our proposed passage representation aggregation approaches.

Considering the baseline methods, \PARADE{Transformer} significantly outperforms the Birch and ELECTRA-MaxP score aggregation approaches for most metrics on both collections.
\PARADE{Transformer}'s ranking effectiveness is comparable with T5-3B on the Robust04 collection while using only $4\%$ of the parameters, though it is worth noting that T5-3B is being used in a zero-shot setting.
CEDR-KNRM and ELECTRA-KNRM, which both use some form of representation aggregation, are significantly worse than \PARADE{Transformer} on title queries and have comparable effectiveness on description queries.
Overall, \PARADE{CNN} and \PARADE{Transformer} are consistently among the most effective approaches, which suggests the importance of performing complex representation aggregation on these datasets.

Results on the Genomics dataset are shown in Table~\ref{tab:genomics}. We first observe that this is a surprisingly challenging task for neural models. Unlike Robust04 and GOV2, where transformer-based models are clearly state-of-the-art, we observe that all of the methods we consider almost always underperform a simple BM25 baseline, and they perform well below the best-performing TREC submission.
It is unclear whether this is due to the specialized domain, the smaller amount of training data, or some other factor. 
Nevertheless, we observe some interesting trends.
First, we see that PARADE approaches can outperform score aggregation baselines. However, we note that statistical significance can be difficult to achieve on this dataset, given the small sample size (64 queries). Next, we notice that \PARADE{Max} performs the best among neural methods. This is in contrast with what we observed on Robust04 and GOV2, and suggests that hierarchically aggregating evidence from different passages is not required on the Genomics dataset.

\subsection{Results on the TREC DL Track and NTCIR WWW-3 Track}\label{sec:dlwww}

We additionally study the effectiveness of PARADE on the TREC DL Track and NTCIR WWW-3 Track.
We report results in this section and refer the readers to the TREC and NTCIR task papers for details on the specific hyperparameters used~\cite{li2020ntcir,li2020trec}.

\begin{table}[tb]
    \centering
        \caption{Ranking effectiveness on TREC DL Track document ranking task.
        PARADE's best result is in bold. 
        The top overall result of of each track is underlined.}
\begin{tabular}{cllcc}\toprule
Year & Group  & Runid                & MAP   & nDCG@10 \\ \hline
\multirow{6}{*}{2019} 
&\multirow{4}{*}{TREC} 
&BM25                 & 0.237 & 0.517   \\
&&ucas\_runid1~\cite{DBLP:conf/trec/ChenLHS19}         & 0.264 & 0.644   \\
&&TUW19-d3-re~\cite{DBLP:conf/trec/HofstatterZH19}          & 0.271 & 0.644   \\
&&idst\_bert\_r1~\cite{DBLP:conf/trec/YanLWBWXS19}       & \underline{0.291} & \underline{0.719}  \\ \cmidrule{2-5}
&\multirow{2}{*}{Ours}
&\PARADE{Max}        & {\bf 0.287} & {\bf 0.679}   \\
&&\PARADE{Transformer} & 0.274 & 0.650   \\ \bottomrule
\multirow{7}{*}{2020} 
&\multirow{5}{*}{TREC} 
&    BM25           & 0.379 & 0.527  \\
&&bcai\_bertb\_docv  & 0.430 & 0.627\\ 
&&fr\_doc\_roberta   & 0.442 & 0.640 \\
&&ICIP\_run1         & 0.433 & 0.662  \\
&&    d\_d2q\_duo    & \underline{0.542} & \underline{0.693} \\\cmidrule{2-5}
&\multirow{2}{*}{Ours}    
& \PARADE{Max}      & \textbf{0.420}   & \textbf{0.613}   \\
&& \PARADE{Transformer}  & 0.403   & 0.601    \\ \bottomrule

\end{tabular}
    \label{tab.DL_Doc}
\end{table}

Results from the TREC Deep Learning Track are shown in Table~\ref{tab.DL_Doc}.
In TREC DL'19, we include comparisons with competitive runs from TREC:
{\tt ucas\_runid1}~\citep{DBLP:conf/trec/ChenLHS19} used BERT-MaxP~\citep{DBLP:conf/sigir/DaiC19} as the reranking method,
{\tt TUW19-d3-re}~\citep{DBLP:conf/trec/HofstatterZH19} is a Transformer-based non-BERT method, and
{\tt idst\_bert\_r1}~\citep{DBLP:conf/trec/YanLWBWXS19} utilizes structBERT~\cite{DBLP:conf/iclr/0225BYWXBPS20}, which is intended to strengthen the modeling of sentence relationships.
All PARADE variants outperform {\tt ucas\_runid1} and {\tt TUW19-d3-re} in terms of nDCG@10, but cannot outperform {\tt idst\_bert\_r1}.
Since this run's pre-trained structBERT model is not publicly available, we are not able to embed it into PARADE and make a fair comparison.
In TREC DL'20, the best TREC run {\tt d\_d2q\_duo} is a T5-3B model.
Moreover, \PARADE{Max} again outperforms \PARADE{Transformer}, which is in line to the Genomics results and in contrast to results on Robust04 and GOV2.
contrast to the previous result in Table~\ref{tb.main_result}.
We explore this further in Section~\ref{sec.bias}.

Results from the NTCIR WWW-3 Track are shown in Table~\ref{tab:run_ntcir}.
{\tt KASYS-E-CO-NEW-1} is a Birch-based method~\cite{DBLP:conf/emnlp/YilmazWYZL19} that uses BERT-Large and {\tt Technion-E-CO-NEW-1} is a cluster-based method.
As shown in Table~\ref{tab:run_ntcir}, \PARADE{Transformer}'s effectiveness is comparable with {\tt KASYS-E-CO-NEW-1} across metrics.
On this benchmark, \PARADE{Transformer} outperforms \PARADE{Max} by a large margin.

\begin{table}[tb]
    \centering
        \caption{Ranking effectiveness of PARADE on NTCIR WWW-3 task. 
        PARADE's best result is in bold. 
        The best result of the Track is underlined.}
    \begin{tabular}{lccc} \toprule
     Model       & nDCG@10  & Q@10   & ERR@10  \\ \hline
     BM25        & 0.5748.  & 0.5850 & 0.6757   \\ 
     Technion-E-CO-NEW-1 & 0.6581 &0.6815 & 0.7791   \\
KASYS-E-CO-NEW-1 & \underline{0.6935}  & \underline{0.7123} & 0.7959  \\
\hline
 \PARADE{Max}    & 0.6337   & 0.6556 & 0.7395   \\
 \PARADE{Transformer}          & {\bf 0.6897}   & {\bf 0.7016} & \underline{{\bf 0.8090}}   \\ \bottomrule
    \end{tabular}
    \label{tab:run_ntcir}
\end{table}

\section{Analysis}
\label{sec:analysis}
\begin{table}[tb] 
\centering
\caption{Comparison with transformers that support longer text sequences on the Robust04 collection.
Baseline results are from~\cite{DBLP:conf/emnlp/JiangXL020}.}
\begin{tabular}{lll} \toprule
Model & nDCG@20  & ERR@20 \\ \midrule
Sparse-Transformer & 0.449 & 0.119 \\
Longformer-QA      & 0.448 & 0.113 \\
Transformer-XH     & 0.450 & 0.123 \\
QDS-Transformer    & 0.457 & 0.126 \\ \midrule
\PARADE{Transformer} & {\bf 0.565} & {\bf 0.149}     \\ \bottomrule
\end{tabular}
\label{table.efficient_trans}
\end{table}

\begin{table*}[bt]
    \centering
    \caption{\PARADE{Transformer}'s effectiveness using BERT models of varying sizes on Robust04 title queries. Significant improvements of distilled over non-distilled models are marked with $\dagger$. ($p<0.01$, two-tailed paired t-test.)} %
        \begin{tabular}{lllllllll} \toprule
  &                      &                             & \multicolumn{2}{c}{Robust04} & \multicolumn{2}{c}{Robust04 (Distilled)} &  Parameter & Inference Time\\
ID                   & Model                 & L / H                            & P@20           & nDCG@20      & P@20           & nDCG@20      & Count  & (ms / doc)   \\ \hline
1                    & BERT-Large           & 24 / 1024        & 0.4508         & 0.5243      &   \textbackslash    &   \textbackslash  & 360M                 & 15.93 \\
2                    & BERT-Base            & 12 / 768         & 0.4486         & 0.5252      &   \textbackslash    &   \textbackslash  & 123M                 & 4.93   \\ \hline
3                    &     \textbackslash   & 10 / 768         & 0.4420         & 0.5168      &   0.4494$^{\dagger}$ &	0.5296$^{\dagger}$         & 109M                 & 4.19\\
4                    &     \textbackslash   & 8 / 768          & 0.4428         & 0.5168      &   0.4490$^{\dagger}$ &	0.5231         & 95M                  & 3.45\\
5                    & BERT-Medium          & 8 / 512          & 0.4303         & 0.5049      &   0.4388$^{\dagger}$	&   0.5110       & 48M                  & 1.94\\
6                    & BERT-Small           & 4 / 512          & 0.4257         & 0.4983      &   0.4365$^{\dagger}$	&   0.5098$^{\dagger}$     & 35M                  & 1.14 \\
7                    & BERT-Mini            & 4 / 256          & 0.3922         & 0.4500      &   0.4046$^{\dagger}$	&   0.4666$^{\dagger}$     & 13M                  & 0.53  \\
8                    &     \textbackslash   & 2 / 512          & 0.4000         & 0.4673      &   0.4038	&   0.4729      & 28M                  & 0.74 \\
9                    & BERT-Tiny            & 2 / 128          & 0.3614         & 0.4216      &   0.3831$^{\dagger}$	&   0.4410$^{\dagger}$    & 5M                   & 0.18  \\ \bottomrule
    \end{tabular}
    \label{tab:model_size}
\end{table*}
In this section, we consider the following research questions:
\begin{itemize}
    \item {\bf RQ1:} How does PARADE perform compare with transformers that support long text?
    \item {\bf RQ2:} How can BERT's efficiency be improved while maintaining its effectiveness?
    \item {\bf RQ3:} How does the number of document passages preserved influence effectiveness?
    \item {\bf RQ4:} When is the representation aggregation approach preferable to score aggregation?
\end{itemize}

\subsection{Comparison with Long-Text Transformers (RQ1)}
Recently, a line of research focuses on reducing the redundant computation cost in the transformer block, allowing models to support longer sequences.
Most approaches design novel sparse attention mechanism for efficiency, which makes it possible to input longer documents as a whole for ad-hoc ranking.
We consider the results reported by~\citet{DBLP:conf/emnlp/JiangXL020} to compare some of these approaches with passage representation aggregation.
The results are shown in Table~\ref{table.efficient_trans}.
In this comparison, long-text transformer approaches achieve similar effectiveness and underperform \PARADE{Transformer} by a large margin.
However, it is worth noting that these approaches use the CLS representation as features for a downstream model rather than using it to predict a relevance score directly, which may contribute to the difference in effectiveness.
A larger study using the various approaches in similar configurations is needed to draw conclusions.
For example, it is possible that QDS-Transformer's effectiveness would increase when trained with maximum score aggregation; this approach could also be combined with PARADE to handle documents longer than Longformer's maximum input length of 2048 tokens.
Our approach is less efficient than that taken by the Longformer family of models, so we consider the question of how to improve PARADE's efficiency in Section~\ref{sec.modelsize}.

\subsection{Reranking Effectiveness vs. Efficiency (RQ2)} \label{sec.modelsize}
While BERT-based models are effective at producing high-quality ranked lists, they are computationally expensive.
However, the reranking task is sensitive to efficiency concerns, because documents must be reranked in real time after the user issues a query.
In this section we consider two strategies for improving PARADE's efficiency.

\paragraphHdTop{Using a Smaller BERT Variant.}
As smaller models require fewer computations, we study the reranking effectiveness of PARADE when using pre-trained BERT models of various sizes, providing guidance for deploying a retrieval system. 
To do so, we use the pre-trained BERT provided by~\citet{DBLP:journals/corr/abs-1908-08962}.
In this analysis we change several hyperparameters to reduce computational requirements: we rerank the top 100 documents from BM25, train with a cross-entropy loss using a single positive or negative document, reduce the passage length 150 tokens, and reduce the stride to 100 tokens.
We additionally use BERT models in place of ELECTRA so that we can consider models with LM distillation
(i.e., distillation using self-supervised PLM objectives), which \citet{gao2020understanding} found to be more effective than ranker distillation alone (i.e., distillation using a supervised ranking objective).
From Table~\ref{tab:model_size}, it can be seen that as the size of models is reduced, their effectiveness decline monotonously.
The hidden layer size (\#6 vs \#7, \#8 vs \#9) plays a more critical role for performance than the number of layers (\#3 vs \#4, \#5 vs \#6).
An example is the comparison between models \#7 and \#8.
Model \#8 performs better; it has fewer layers but contains more parameters.
The number of parameters and inference time are also given in Table~\ref{tab:model_size} to facilitate the study of trade-offs between model  complexity and effectiveness.

\paragraphHd{Distilling Knowledge from a Large Model.}
To further explore the limits of smaller PARADE models, we apply knowledge distillation to leverage knowledge from a large teacher model.
We use \PARADE{Transformer} trained with BERT-Base on the target collection as the teacher model.
Smaller student models then learn from the teacher at the output level.
We use mean squared error as the distilling objective, which has been shown to work effectively~\cite{DBLP:journals/corr/abs-2004-11045,DBLP:journals/corr/abs-1903-12136}.
The learning objective penalizes the student model based on both the ground-truth and the teacher model:
\begin{align}\label{eq:loss_kd}
    L & = \alpha \cdot L_{CE} + (1 - \alpha) \cdot ||z^t - z^s||^2
\end{align}
where $L_{CE}$ is the cross-entropy loss with regard to the logit of the student model and the ground truth, $\alpha$
weights the importance of the learning objectives, and
$z^t$ and $z^s$ are logits from the teacher model and student model, respectively.

As shown in Table~\ref{tab:model_size}, the nDCG@20 of distilled models always increases.
The PARADE model using 8 layers (\#4) can achieve comparable results with the teacher model.
Moreover, the PARADE model using 10 layers (\#3) can  outperform the teacher model with 11\% fewer parameters.
The PARADE model trained with BERT-Small achieves a nDCG@20 above 0.5, which outperforms BERT-MaxP using BERT-Base, while
requiring only 1.14 ms to perform inference on one document.
Thus, when reranking 100 documents, the inference time for each query is approximately 0.114 seconds.

\subsection{Number of Passages Considered (RQ3)} \label{sec.numPassages}

\begin{figure}[tb]
    \centering
    \includegraphics[ height=2.2in]{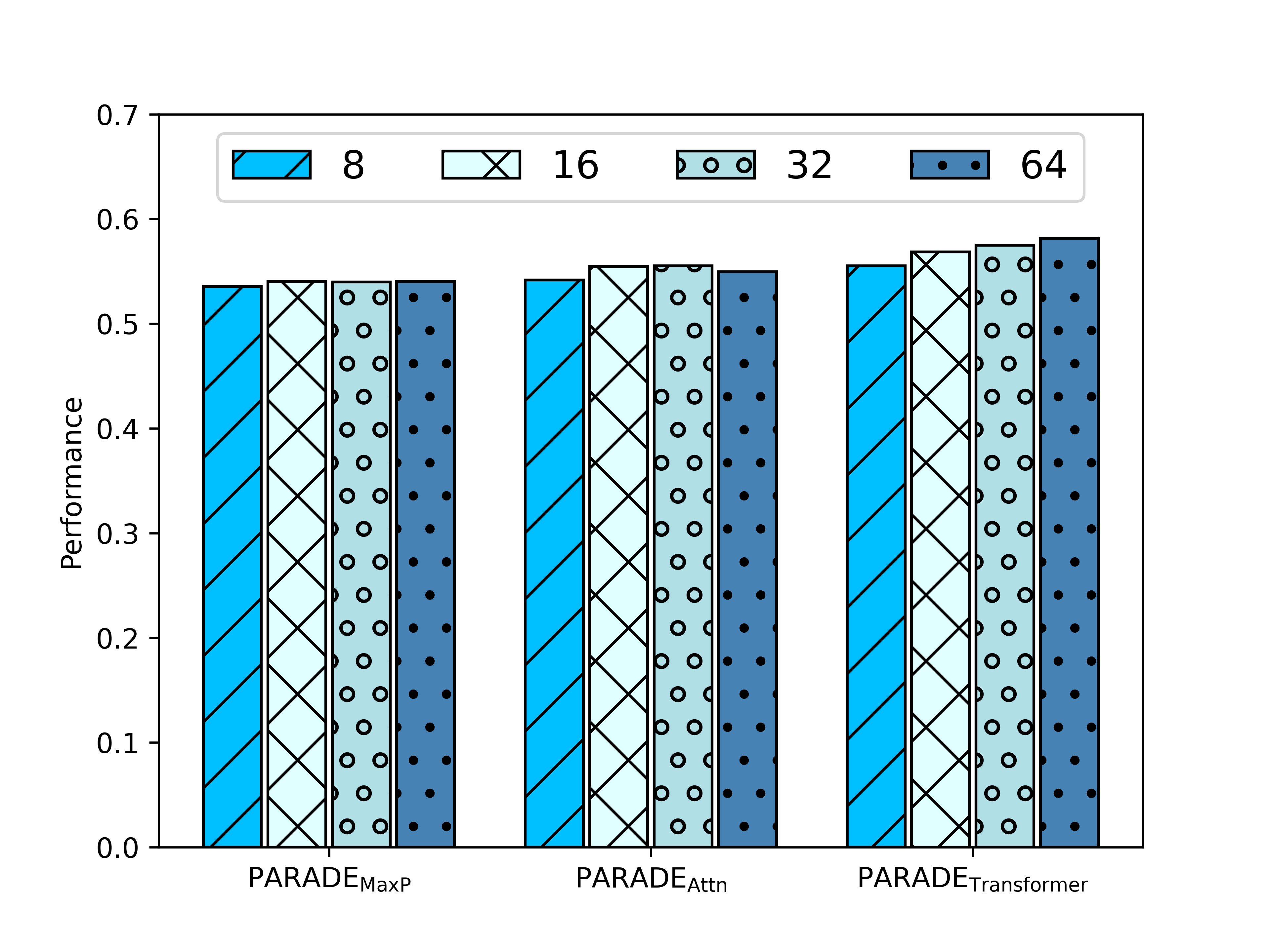}
    \caption{Reranking effectiveness of \PARADE{Transformer} when different number of passages are being used on {\it Gov2} title dataset. nDCG@20 is reported.}
    \label{fig.parameter_sensitivity}
\end{figure}

\begin{table}[tb]
    \centering
     \caption{Reranking effectiveness of \PARADE{Transformer} using various preserved data size on {\it GOV2} title dataset. 
    nDCG@20 is reported.
    The indexes of columns and rows are number of passages being used.}
    \begin{tabular}{ccccc} \toprule
Train \textbackslash \space Eval 
& 8  & 16   & 32  & 64   \\ \hline
8   & {\it 0.5554} & 0.5648 & 0.5648 & 0.5680  \\
16  & 0.5621 & {\it 0.5685} & 0.5736 & 0.5733 \\
32  & 0.5610 & 0.5735 & {\it 0.5750} & 0.5802 \\
64  & 0.5577 & 0.5665 & 0.5760 & {\it 0.5815} \\ \bottomrule
    \end{tabular}
   
    \label{tab:train_eval}
\end{table}

One hyper-parameter in PARADE is the maximum number of passages being used, i.e., preserved data size, which is studied to answer RQ3 in this section.
We consider title queries on the GOV2 dataset given that these documents are longer on average than in Robust04.
We use the same hyperparameters as in Section~\ref{sec.modelsize}.
Figure~\ref{fig.parameter_sensitivity} depicts nDCG@20 of \PARADE{Transformer} with the number of passages varying from 8 to 64.
Generally, larger preserved data size results in better performance for \PARADE{Transformer}, which suggests that a document can be better understood from document-level context with more preservation of its content.
For \PARADE{Max} and \PARADE{Attn}, however, the performance degrades a little when using 64 passages.
Both max pooling (Max) and simple attention mechanism (Attn) have limited capacity and are challenged when dealing with such longer documents.
The \PARADE{Transformer} model is able to improve nDCG@20 as the number of passages increases, demonstrating its superiority in detecting relevance when documents become much longer.

However, considering more passages also increases the number of computations performed.
One advantage of the PARADE models is that the number of parameters remains constant as the number of passages in a document varies.
Thus, we consider the impact of varying the number of passages considered between training and inference.
As shown in Table~\ref{tab:train_eval}, rows indicate the number of passages considered at training time while columns indicate the number used to perform inference.
The diagonal indicates that preserving more of the passages in a document consistently improves nDCG.
Similarly, increasing the number of passages considered at inference time (columns) or at training time (rows) usually improves nDCG.
In conclusion, the number of passages considered plays a crucial role in PARADE's effectiveness.
When trading off efficiency for effectiveness, PARADE models' effectiveness can be improved by training on more passages than will be used at inference time. This generally yields a small nDCG increase.

\subsection{When is the representation aggregation approach preferable to score aggregation? (RQ4)} \label{sec.bias}
While PARADE variants are effective across a range of datasets and the \PARADE{Transformer} variant is generally the most effective, this is not always the case.
In particular, \PARADE{Max} outperforms \PARADE{Transformer} on both years of TREC DL and on TREC Genomics.
We hypothesize that this difference in effectiveness is a result of the focused nature of queries in both collections.
Such queries may result in a lower number of highly relevant passages per document, which would reduce the advantage of using more complex aggregation methods like \PARADE{Transformer} and \PARADE{CNN}.
This theory is supported by the fact that TREC DL shares queries and other similarities with MS MARCO, which only has 1--2 relevant passages per document by nature of its construction.
This query overlap suggests that the queries in both TREC DL collections \textit{can} be sufficiently answered by a single highly relevant passage.
However, unlike the shallow labels in MS MARCO, documents in the DL collections contains deep relevance labels from NIST assessors.
It is unclear how often documents in DL also have only a few relevant passages per document.

We test this hypothesis by using passage-level relevance judgments to compare the number of highly relevant passages per document in various collections.
To do so, we use mappings between relevant passages and documents for those collections with passage-level judgments available: TREC DL, TREC Genomics, and GOV2.
We create a mapping between the MS MARCO document and passage collections by using the MS MARCO Question Answering (QnA) collection to map passages to document URLs.
This mapping can then be used to map between passage and document judgments in DL'19 and DL'20. %
With DL'19, we additionally use the FIRA passage relevance judgments~\cite{DBLP:conf/cikm/HofstatterZSSH20} to map between documents and passages.
The FIRA judgments were created by asking annotators to identify relevant passages in every DL'19 document with a relevance label of 2 or 3 (i.e., the two highest labels).
Our mapping covers nearly the entire MS MARCO collection, but it is limited by the fact that DL's passage-level relevance judgments may not be complete.
The FIRA mapping covers only highly-relevant DL'19 documents, but the passage annotations are complete and it was created by human annotators with quality control.
In the case of TREC Genomics, we use the mapping provided by TREC.
For GOV2, we use the sentence-level relevance judgments available in WebAP~\cite{keikha2014evaluating,keikha2014retrieving}, which cover 82 queries. %

We compare passage judgments across collections by using each collection's annotation guidelines to align their relevance labels with MS MARCO's definition of a relevant passage as one that is \textit{sufficient} to answer the question query.
With GOV2 we consider passages with a relevance label of 3 or 4 to be relevant.
With DL documents we consider a label of 2 or 3 to be relevant and passages with a label of 3 to be relevant.
With FIRA we consider label 3 to be relevant.
With Genomics we consider labels 1 or 2 to be relevant.

We align the maximum passage lengths in GOV2 to FIRA's maximum length so that they can be directly compared.
To do so, we convert GOV2's sentence judgments to passage judgments by collapsing sentences following a relevant sentence into a single passage with a maximum passage length of 130 tokens, as used by FIRA\footnote{Applying the same procedure to both FIRA and WebAP with longer maximum lengths did not substantially change the trend.}.
We note that this process can only \textit{decrease} the number of relevant passages per document observed in GOV2, which we expect to have the highest number.
With the DL collections using the MS MARCO mapping, the passages are much smaller than these lengths, so collapsing passages could only \textit{decrease} the number of relevant passages per document.
We note that Genomics contains ``natural'' passages that can be longer; this should be considered when drawing conclusions.
In all cases, the relevant passages comprise a small fraction of the document.

In each collection, we calculate the number of relevant passages per document using the collection's associated document and passage judgments.
The results are shown in Table~\ref{tab:analysis}.
First, considering the GOV2 and MS MARCO collections that we expect to lie at opposite ends of the spectrum, we see that 38\% of GOV2 documents contain a single relevant passage, whereas 98--99\% of MS MARCO documents contain a single relevant passage.
This confirms that MS MARCO documents contain only 1--2 highly relevant passages per document by nature of the collection's construction.
The percentages are the lowest on GOV2 as expected.
While we would prefer to put these percentages in the context of another collection like Robust04, the lack of passage-level judgments on such collections prevents us from doing so.
Second, considering the Deep Learning collections, we see that DL'19 and DL'20 exhibit similar trends regardless of whether our mapping or the FIRA mapping is used.
In these collections, the majority of documents contain a single relevant passage and the vast majority of documents contain one or two relevant passages.
We call this a ``maximum passage bias.''
The fact that the queries are shared with MS MARCO likely contributes to this observation, since we know the vast majority of MS MARCO question queries can be answered by a single passage.
Third, considering Genomics 2006, we see that this collection is similar to the DL collections.
The majority of documents contain only one relevant passage, and the vast majority contain one or two relevant passages.
Thus, this analysis supports our hypothesis that the difference in \PARADE{Transformer}'s effectiveness across collections is related to the number of relevant passages per document in these collections.
\PARADE{Max} performs better when the number is low, which may reflect the reduced importance of aggregating relevance signals across passages on these collections.

\begin{table}[t]
    \centering
    \caption{Percentage of documents with a given number of relevant passages.}
    \scriptsize
    \begin{tabular}{r@{\hskip 0.2in}c@{\hskip 0.1in}c@{\hskip 0.1in}c@{\hskip 0.1in}c@{\hskip 0.1in}c@{\hskip 0.05in}c}
        \toprule
        \textbf{\# Relevant} & GOV2 & DL19 & DL19 & DL20 & MS MARCO & Genomics \\
        \textbf{Passages} & & (FIRA) & (Ours) & (Ours) & train / dev  & 2006 \\
        \midrule
        1 & 38\% & 66\% & 66\% & 67\% & 99\% / 98\%  & 62\% \\
        1--2 & 60\% & 87\% & 86\% & 81\% & 100\% / 100\% & 80\% \\
        3+ & 40\% & 13\% & 14\% & 19\% & 0\% / 0\%  & 20\% \\
         \bottomrule
    \end{tabular}
    \label{tab:analysis}
\end{table}

\section{Conclusion} \label{sec:conclusion}
We proposed the PARADE end-to-end document reranking model
and demonstrated its effectiveness on ad-hoc benchmark collections.
Our results indicate the importance of incorporating diverse relevance signals from the full text into ad-hoc ranking, rather than basing it on a single passage.
We additionally investigated how model size affects performance, finding that
knowledge distillation on PARADE boosts the performance of smaller PARADE models while substantially reducing their parameters. 
Finally, we analyzed dataset characteristics are to explore when representation aggregation strategies are more effective.

\begin{acks}
This work was supported in part by Google Cloud and the TensorFlow Research Cloud.
\end{acks}

\bibliographystyle{ACM-Reference-Format}
\bibliography{8-Reference}

\newpage
\newpage 
\appendix
\section{Appendix}
\label{sec:appendix}

\subsection{Results on the TREC-COVID Challenge}
\label{sec:covid}
\begin{table}[tbh]
    \centering
    \resizebox{.5\textwidth}{!}{
\begin{tabular}{llllll}
\toprule
& runid              & nDCG@10 & P@5    & bpref  & MAP    \\ \hline
1 & {\bf mpiid5\_run3}       & 0.6893  & 0.8514 & 0.5679 & 0.3380 \\
2 & {\bf mpiid5\_run2}       & 0.6864  & 0.8057 & 0.4943 & 0.3185 \\
3 & SparseDenseSciBert & 0.6772  & 0.7600 & 0.5096 & 0.3115 \\
4 & {\bf mpiid5\_run1}       & 0.6677  & 0.7771 & 0.4609 & 0.2946 \\
5 & UIowaS\_Run3       & 0.6382  & 0.7657 & 0.4867 & 0.2845 \\
\bottomrule
\end{tabular}}
    \caption{Ranking  effectiveness of different retrieval systems in the TREC-COVID Round 2.}
    \label{tab:covid_r2}
\end{table}

\begin{table}[htb]
    \centering
        \resizebox{.5\textwidth}{!}{
\begin{tabular}{llllll}
\toprule
& runid              & nDCG@10 & P@5    & bpref  & MAP    \\ \hline
1 & covidex.r3.t5\_lr      & 0.7740  & 0.8600 & 0.5543 & 0.3333 \\
2 & BioInfo-run1           & 0.7715  & 0.8650 & 0.5560 & 0.3188 \\
3 & UIowaS\_Rd3Borda       & 0.7658  & 0.8900 & 0.5778 & 0.3207 \\
4 & udel\_fang\_lambdarank & 0.7567  & 0.8900 & 0.5764 & 0.3238 \\
\hdashline
11 & sparse-dense-SBrr-2    & 0.7272  & 0.8000 & 0.5419 & 0.3134  \\
13 & {\bf mpiid5\_run2}          & 0.7235  & 0.8300 & 0.5947 & 0.3193 \\
16 & {\bf mpiid5\_run1} (Fusion)          & 0.7060  & 0.7800 & 0.6084 & 0.3010 \\
43 & {\bf mpiid5\_run3} (Attn)           & 0.3583  & 0.4250 & 0.5935 & 0.2317 \\
\bottomrule
    \end{tabular}}
    \caption{Ranking  effectivenes of different retrieval systems in the TREC-COVID Round 3.}
    \label{tab:covid_r3}
\end{table}

\begin{table}[htb]
    \centering
    \resizebox{.5\textwidth}{!}{
\begin{tabular}{llllll}
\toprule
& runid                   & nDCG@20 & P@20   & bpref  & MAP    \\ \hline
1 & UPrrf38rrf3-r4          & 0.7843  & 0.8211 & 0.6801 & 0.4681 \\
2 & covidex.r4.duot5.lr     & 0.7745  & 0.7967 & 0.5825 & 0.3846 \\
3 & UPrrf38rrf3v2-r4 & 0.7706 & 0.7856 & 0.6514 & 0.4310 \\
4 & udel\_fang\_lambdarank  & 0.7534  & 0.7844 & 0.6161 & 0.3907 \\
5 & run2\_Crf\_A\_SciB\_MAP & 0.7470  & 0.7700 & 0.6292 & 0.4079 \\
6 & run1\_C\_A\_SciB & 0.7420 & 0.7633 & 0.6256 & 0.3992 \\
7 & {\bf mpiid5\_run1}            & 0.7391  & 0.7589 & 0.6132 & 0.3993 \\ \bottomrule
\end{tabular}}
    \caption{Ranking  effectiveness of different retrieval systems in the TREC-COVID Round 4.}
    \label{tab:covid_r4}
\end{table}
In response to the urgent demand for reliable and accurate retrieval of COVID-19 academic literature, TREC has been developing the TREC-COVID challenge to build a test collection during the pandemic~\cite{DBLP:journals/corr/abs-2005-04474}.
The challenge uses the CORD-19 data set~\cite{DBLP:journals/corr/abs-2004-10706}, which is a dynamic collection enlarged over time.
There are supposed to be 5 rounds for the researchers to iterate their systems.
TREC develops a set of COVID-19 related topics, including queries (key-word based), questions, and narratives.
A retrieval system is supposed to generate a ranking list corresponding to these queries.

We began submitting PARADE runs to TREC-COVID from Round 2.
By using PARADE, we are able to utilize the full-text of the COVID-19 academic papers.
We used the question topics since it works much better than other types of topics.
In all rounds, we employ the \PARADE{Transformer} model.
In Round 3, we additionally tested \PARADE{Attn} and a combination of \PARADE{Transformer} and \PARADE{Attn} using reciprocal rank fusion~\cite{10.1145/1571941.1572114}.

Results from TREC-COVID Rounds 2-4 are shown in Table~\ref{tab:covid_r2}, Table~\ref{tab:covid_r3}, and Table~\ref{tab:covid_r4}, respectively.\footnote{Further details and system descriptions can be found at \url{https://ir.nist.gov/covidSubmit/archive.html}}
In Round 2, PARADE achieves the highest nDCG, further supporting its effectiveness.\footnote{To clarify, the run type of the PARADE runs is feedback, but they were cautiously marked as manual due to the fact that they rerank a first-stage retrieval approach based on {\tt udel\_fang\_run3}. Many participants did not regard this as sufficient to change a run's type to manual, however, and the PARADE runs would be regarded as feedback runs following this consensus.}
In Round 3, our runs are not as competitive as the previous round.
One possible reason is that the collection doubles from Round 2 to Round 3, which can introduce more inconsistencies between training and testing data as we trained PARADE on Round 2 data and tested on Round 3 data.
In particular, our run {\tt mpiid5\_run3} performed poorly.
We found that it tends to retrieve more documents that are not likely to be included in the judgment pool.
When considering the bpref metric that takes only the judged documents into account, its performance is comparable to that of the other variants.
As measured by nDCG, PARADE's performance improved in Round 4 (Table~\ref{tab:covid_r4}), but is again outperformed by other approaches.
It is worth noting that the PARADE runs were created by single models (excluding the fusion run from Round 3), whereas e.g. the {\tt UPrrf38rrf3-r4} run in Round 4 is an ensemble of more than 20 runs.

\end{document}